\documentstyle[twoside,fleqn,espcrc2,psfig]{article}

\newcommand{\AmS}{{\protect\the\textfont2
   A\kern-.1667em\lower.5ex\hbox{M}\kern-.125emS}}

\newcommand{\gtp}{{\tilde g}'}
\newcommand{\sint}{\sin^2{\theta_W}}

\newcommand{\Mz}{{M_z}}

\title{A Half-Composite Standard Model at a TeV and $\sint$}

\author{Savas Dimopoulos\address{Physics Department, Stanford
University, Stanford, CA 94305-4060, USA.} and David
Elazzar Kaplan\address{SLAC, Stanford, CA 94025}
          }

\begin{document}

\begin{abstract}

We apply a recently proposed mechanism -- in which an $SU(3)$ symmetry
predicts the weak mixing angle -- to construct realistic theories with 
composite quarks and leptons at a TeV.
Although the models are strongly coupled, they are reliably analyzed using 
complementarity and 't Hooft's anomaly matching.  In the simplest models 
the right-handed fermions are composite, while the left-handed are 
elementary. Strong $SU(2)_R$ forces give rise to 12-particle 
instanton-mediated processes. They violate baryon and lepton numbers by 
three units and result in spectacular multilepton and multijet events 
at the LHC.  Models in which the leptons are in an $SU(3)$-triplet
can be directly tested in muonium-antimuonium conversion experiments.

\vspace{1pc}
\end{abstract}

\maketitle
\section{Introduction}
The one quantitative success of physics beyond the standard model (SM) is
the prediction of the weak mixing angle
by supersymmetric grand unified theories (GUTs)
\cite{Dimopoulos:1981zb}.  Running the $SU(5)$ prediction  $\sint=3/8$
  \cite{Georgi:1974sy,Georgi:1974yf} from the GUT scale
to the weak scale in supersymmetric theories
produces the measured value of $0.231$ within theoretical uncertainties.
This requires the existence of a
large energy desert above the weak scale.  For a theory with a low cutoff,
a different approach is necessary.

Recently, a new mechanism for predicting the weak mixing angle with 
TeV-physics was proposed \cite{Dimopoulos:2002mv}.
It leads to the unification of the two electroweak gauge couplings into 
their $SU(3)$-symmetric value, giving
$\sint=.25$ at tree level. Since this value is close to the experimental 
value of $\sint=.231$ at $M_Z$,
``SU(3)-unification'' occurs at a few TeV \cite{Dimopoulos:2002mv}. 

In this paper we show how this mechanism can be incorporated in models with 
composite quarks and leptons. Specifically,
we consider models with an $SU(2)_R$ gauge group that becomes strong and 
confines at around a TeV. The resulting
composite fermions are all the right handed quarks and leptons of the 
standard model. All the models predict the correct
value of  $\sint$ to within a few percent, provided that the compositeness 
scale is near a TeV  -- as motivated by the
hierarchy problem. In general, it is difficult to analyze the spectrum 
of strongly coupled field theories, let
alone predict any quantities to within a few percent. Yet, in the models we 
propose all the properties of interest can
be reliably analyzed. One reason is that the strongly coupled phase of our 
models is ``complementary'' to the weakly
coupled Higgs phase which can be studied within perturbation theory 
\cite{Dimopoulos:1980hn}.

The crucial ingredient guaranteeing complementarity is that all the Higgs 
fields that develop VEVs are in the fundamental representation 
of the gauge group. As a 
result, the Wilson line is always screened by the
Higgs field and does not discriminate the confinement and Higgs phases -- 
leading to the conjecture that there is no
essential difference between the two \cite{tHooft,Fradkin:1979dv}. 
In particular, the spectrum of 
massless composite fermions produced by the strong
confinement dynamics is deduced by the much easier task of reading-off the 
massless fermions in the Higgs phase. An
important check of this strategy is 't Hooft's anomaly matching condition. 
It states that any exact global 
symmetry spontaneously unbroken by the strong confinement dynamics, must 
have the same global anomalies in the infrared as it
does in the ultraviolet. In others words, the anomalies of the of the 
unbroken global currents must be the same weather
evaluated by using the elementary fermions or the massless composite 
fermions produced by the strong dynamics.
Complementarity is a good strategy for looking for theories satisfying 
't Hooft's matching conditions. We will use both
of these tools to ensure that we indeed get the spectrum of the standard 
model in our theories.

In the section 2 we will discuss a minimal realistic composite model 
predicting $\sint$. In section 3 we will
generalize it and we will conclude with experimental signatures in section 4.

\section{The Strong Left-Right Model}
Before we describe the model, we review the mechanism which
predicts the correct low energy value of $\sint$ with relatively
small theoretical uncertainties.

Start with the SM, $SU(3)_c \times SU(2) \times U(1)$ with
all matter and Higgs in their normal representations, and add
an $SU(3)_W$ gauge symmetry.  
A single scalar representation $\Sigma$ connects the $SU(3)_W$
to $SU(2)\times U(1)$ -- it is a triplet of $SU(3)_W$ and 
has the Higgs quantum numbers $(2,+1/2)$ under the $SU(2)\times U(1)$.  
A generic potential allows $\Sigma$ to get a VEV which breaks 
$SU(3)_W \times SU(2) \times U(1)$ to the diagonal 
$SU(2)_L \times U(1)_Y$.  If the gauge couplings of
the original $SU(2) \times U(1)$ are at least somewhat bigger than
the $SU(3)_W$ coupling, then there is a large region in parameter
space in which the low energy couplings reflect the $SU(3)$
symmetry up to a few percent correction \cite{Dimopoulos:2002mv}.
The symmetry insures that at the breaking scale $M$, the low energy 
gauge couplings approximately satisfy the $SU(3)$ relation 
$g/g' = \sqrt{3}$ or $\sint=.25$ 
\cite{Weinberg:1972nd}.

Now we gauge $SU(2)_R$.
The full gauge sector of the model contains an $SU(3)_W \times
SU(3)_c \times SU(2)_{\tilde L} \times SU(2)_R \times U(1)_X$ symmetry.
The fermion content consists of three generations of quarks
$q: (1,3,2,1,+1/6)$ and $q^c:(1,{\bar 3},1,2,-1/6)$ and leptons
$\ell: (1,1,2,1,-1/2)$ and $\ell^c: (1,1,1,2,+1/2)$, 
which become the standard model content
plus right handed neutrinos after left-right breaking.  

We now describe the scalars and their roles in the breaking
to the standard model when the theory is in the Higgs phase.  
First, it includes $\phi: (1,1,1,2,+1/2)$, responsible 
for breaking $SU(2)_R \times U(1)_X \rightarrow U(1)_{\tilde Y}$.
A field $\Sigma: (3,1,2,2,0)$ is responsible for breaking $SU(3)_W 
\times SU(2)_{\tilde L} \times U(1)_{\tilde Y} \rightarrow SU(2)_L \times U(1)_Y$, 
the electroweak sector of the standard model.  This breaking results 
in the value $\sint \simeq .25$ when the Left-Right and $x$ couplings are 
somewhat large \cite{Dimopoulos:2002mv}.  Finally, the standard model 
Higgs field is
contained in $h: (1,1,2,2,0)$.

Next we switch to the confinement picture by taking the $SU(2)_R$ 
gauge coupling strong. 
Since all scalars are in the fundamental
representation of $SU(2)_R$ we expect complementarity between the
Higgs and confining descriptions to hold.  Physics below the confining
scale looks like the standard model; this matches the physics of the
Higgs picture below the breaking scale.  

\noindent{\bf 't Hooft Anomaly Matching:}
While the tools
of complementarity allow us to construct the theory in the infrared, we
perform a non-trivial check that the low energy spectrum is correct via
't Hooft anomaly matching \cite{tHooft}.  To do so, we find all of the global
anomalies in the ultraviolet (UV) and infrared (IR) and see if they are 
equal for the symmetry not spontaneously broken by the strong dynamics.

When the $SU(3)_c \times SU(2)_{\tilde L} \times U(1)_X$ couplings and all 
Yukawa couplings are turned off, there exists a global $SU(12)_f$ 
symmetry which transforms the 12 fermion $SU(2)_R$ doublets -- three quarks 
and one lepton per generation -- as the fundamental 
representation\footnote{Classically the symmetry is $U(12)_f$ but 
the $U(1)$ is anomalous.}.  
This symmetry has a global $SU(12)_f^3$ anomaly with two ${\bf 12}\,$s
contributing and no ${\bf {\overline 12}}\,$s.

Turning off the gauge coupling of the 
$SU(3)_W$ along with any scalar potential couplings leaves a global $U(9)_s$ 
symmetry under which $\Sigma$, $\phi$ and $h$ together form a nine-dimensional
representation.  We will number the $SU(2)_R$ doublets in these scalars
as follows:
\begin{eqnarray}
\Sigma & = & \bordermatrix{ 
	& {\bf 2_L} & \rightarrow \cr
	\uparrow & 1 & 2 \cr
	{\bf 3_W} & 3 & 4 \cr
	\downarrow & 5 & 6 \cr}\nonumber\\
h & = & \bordermatrix{
	&\cr
	{\bf 2_L} & 7 \cr
	\downarrow & 8 \cr}\\\nonumber
\phi & = & (9)
\end{eqnarray}
Because these global symmetries only rotate scalars, there are
trivially no additional global anomalies beyond $SU(12)_f^3$ mentioned
above.

Thus the full global symmetry of the theory
at high energies is $SU(12)_f\times SU(9)_s\times U(1)_s$ and the matter content
is the fermion doublets $\{q^c,\ell^c\} \equiv \psi^a$ in the fundamental 
representation of $SU(12)_f$ and scalar doublets $\{\Sigma, h,\phi\} \equiv H^i$
in the fundamental representation of $SU(9)_s$ and carrying a unit of $U(1)_s$
charge.

We postulate the existence of a pair of condensates due to the 
$SU(2)_R$ gauge coupling getting strong which breaks the weakly 
coupled gauge groups $SU(3)_W \times SU(2)_{\tilde L} \times U(1)_X$ to the 
electroweak group $SU(2)_L \times U(1)_Y$.
This breaking is accomplished by
\begin{eqnarray}
\langle H^{\dagger} H \rangle &\sim&
	\bordermatrix{ & 1 & \phantom{0} & 4 & \phantom{0} &\phantom{0} & 9 \cr
	1 & M^2 & & M^2 & & & \cr
	\phantom{0} & & & & & & \cr
	4 & M^2 & & M^2 & & & \cr
	\phantom{0} & & & & & & \cr
	\phantom{0} & & & & & & \cr
        9 & & & & & & {\cal V}^2 }
\label{condensate1}\\
\langle H \epsilon H \rangle &\sim& 
	\bordermatrix{ & 1 & \phantom{0} & 4 & \phantom{0} & 9 \cr
	1 & & & & & - M {\cal V} \cr
	\phantom{0} & & & & & \cr
	4 & & & & & - M {\cal V} \cr
	\phantom{0} & & & & & \cr
	\phantom{0} & & & & & \cr
        9 & M {\cal V} & & M {\cal V} & & }
\label{condensate2}
\end{eqnarray}
where the $SU(2)_R$ indices are contracted and $M\sim {\cal V}$.
This is the vev that would arise by plugging in the vev for $\Sigma$ and
$\phi$ from the weakly-coupled version of this story.  

The first of the condensates above break $SU(9)_s\times U(1)_s$ to 
$SU(7)_s\times U(1)_s^3$.  The second breaks one linear combination
of the $U(1)$s so the remaining global symmetry group is
$SU(7)_s\times U(1)_s^2$.  The low energy theory will contain the light
scalars which are the goldstone bosons as a result of the symmetry
breaking (though some will be eaten since a portion of these symmetries
are gauged), and light fermions which will reproduce the global anomalies
of the UV theory as required by anomaly matching 
\cite{tHooft,Coleman:1982yg,Frishman:1981dq}.

In the IR we need to find composite fermions which
produce the same anomalies in the global symmetries as those in
the UV theory.  The simplest set of composites, and their representations
under the global $SU(12)_s\times SU(7)_s \times U(1)_s^2$
symmetry are
\begin{eqnarray}
(H^i \epsilon \psi^a): &\:\:& (12,1,a,b) \nonumber\\
			&\: +& (12,7,0,c) + (12,1,-a,-b)\nonumber\\
(H_i^{\dagger} \psi^a): &\:\:& (12,1,-a,-b) \nonumber\\
			&\: +& (12,{\bar 7},0,c) + (12,1,a,b)
\label{composites}
\end{eqnarray}
where $a,b,c\neq 0$.  

The $U(1)$ charges can be understood as follows:
The two unbroken $U(1)$s must act trivially on the non-zero elements
in (\ref{condensate2}).  If we diagonalize the matrix in 
eq. (\ref{condensate1}) using $SU(9)_s$ transformations making only the
(1,1) and (9,9) entries non-zero, the other condensate (\ref{condensate2}) 
under the same rotation becomes:
\begin{eqnarray}
\langle H \epsilon H \rangle &\sim&
	\bordermatrix{ & 1 & \phantom{0} & \phantom{0} & 9 \cr
	1 & & & & - M {\cal V} \cr
	\phantom{0} & & & & \cr
	\phantom{0} & & & & \cr
	\phantom{0} & & & & \cr
	\phantom{0} & & & & \cr
        9 & M {\cal V} & & & }
\end{eqnarray}
Thus the unbroken $U(1)_s$ charges of the first and ninth components of
$H$ (the $SU(7)_s$ singlets) must be equal and opposite.  
One $U(1)_s$ can be taken as a subgroup of the full $SU(9)_s$, 
and is therefore traceless.  There exists a basis for the two $U(1)$s
such that the charges chosen in eq. (\ref{composites}) are the most 
general set.

Now, to reproduce the $SU(12)_s^3$ anomaly of the UV theory, we need two
and only two 12-dimensional representations of fermions.  Therefore,
anomaly matching requires that only the $SU(7)_s$ singlets
can be the chiral matter in the low-energy theory. 
The remaining global anomalies must all vanish.  
This is accomplished by choosing two ${\bf 12}\,$s which are vector-like 
under the remaining quantum numbers.  The choice is then just
one of each linear combination of $SU(7)_s$ singlets with the same
quantum numbers.

Finally, we need to indentify the quantum numbers of the composite
fermions under the unbroken gauge symmetries.  Since the unbroken
$SU(2)_L$ acts trivially on the original UV fermions $\psi^a$, then
the composite fermions which are $SU(7)_s$ singlets will be $SU(2)_L$ 
singlets.  The unbroken $U(1)_Y$ is a direct sum of the $U(1)_X$
generator and the $T^8$ generator of $SU(3)_W$ with the normalization
$T^8 = diag\{-1/2,-1/2,+1\}$ as it acts on triplets.  The action
of $U(1)_Y$ on the fermions $\psi$ and scalars $H$ is via the following 
representations of the generator, respectively:
\begin{eqnarray*}
&diag\{(-\frac{1}{6},-\frac{1}{6},-\frac{1}{6},\frac{1}{2}),(\cdots),(\cdots)\}\\
&diag\{-\frac{1}{2},-\frac{1}{2},1,-\frac{1}{2},-\frac{1}{2},1,0,0,\frac{1}{2}\}
\end{eqnarray*}
where we construct the fermion multiplet as 
$\psi = \{q_1^{\alpha_1},\ell_1,q_2^{\alpha_2},\ell_2,q_3^{\alpha_3},\ell_3\}$;
the subscripts indicate generation number and the $\alpha_i$ are color indicies.
We see that the $SU(7)_s$-singlet parts of $H$ (the first and last 
components) have hypercharges $\pm 1/2$ and thus the composite
quarks have hypercharges $-1/2 - 1/6 = -2/3$ and $1/2 - 1/6 = 1/3$
and composite leptons have hypercharges $-1/2 + 1/2 = 0$ and 
$1/2 + 1/2 = 1$ giving three generations of $u^c$, $d^c$, $\nu_R^c$
and $e^c$ respectively.  The low energy theory has exactly the fermion
content of the SM plus three generations of right-handed neutrinos.

\noindent{\bf Fermion Masses:}
The light scalars will be the goldstone bosons associated with the
breaking of approximate global symmetries.  They can be defined as
the variations around the condenstates (\ref{condensate1}) and (\ref{condensate2}).
Nearly all of them can be described as variations around the first condensate:
\begin{equation}
\delta\langle H^{\dagger} H \rangle \equiv U = e^{i \pi/M}
\end{equation}
where $\pi = \pi^a t^a$ is sum over only the broken generators of the original
$SU(9)_s$ global symmetry and contains two complex {\bf 7} representations of
$SU(7)_s$ and a complex singlet.  The {\bf 7} contains the Higgs and components
of $\Sigma$, some of which are eaten by $SU(3)_W$ breaking.

Without Yukawa couplings in the UV theory, Yukawas do not appear in the IR and
an exact chiral symmetry leaves the fermions massless.  If we introduce
Yukawa couplings $\lambda^u q h q^c$ and $\lambda^d q h^{\dagger} q^c$ in
the UV theory, we can treat $\lambda^q q$ as spurions for global symmetry
breaking.  The global-symmetry-invariant operator becomes 
$(\lambda^q q) H \psi$, where $(\lambda^q q)$ transforms as a 
$({\bf {\bar 12}},{\bf {\bar 9}},-1)$ under 
$SU(12)_s \times SU(9)_s \times U(1)_s$.
Using na\"ive dimensional analysis (NDA) 
\cite{Weinberg:1979kz,Manohar:1984md,weak},
we estimate the size of the operators to be
\begin{equation}
\frac{\Lambda^4}{g^2} \left[ \frac{\lambda^q q}{\Lambda^{3/2}} U \frac{g \Psi}{\Lambda^{3/2}} \right]
\end{equation}
where we take $\Lambda/g = \Lambda/4\pi = f_M \equiv M$.  Thus when $U$ is
expanded to first order, NDA gives a Yukawa strength in the IR equal to
that in the UV -- a result which agrees with what one would have 
expected from complementarity.

The theory is a two-Higgs-doublet version of the
SM.  If only the ``up-type'' Higgs gets a VEV, then the couplings
$\lambda^u q h q^c$, $\lambda^d q h^{\dagger} q^c$ and 
$\lambda^e \ell h^{\dagger} \ell^c$ in the UV theory produce the
necessary charged fermion masses in the low energy theory.  The neutrino
masses would come from the coupling $\lambda^{\nu} \ell h \ell^c$ thus
requiring $\lambda^{\nu}$ to be extremely small.  This can be ameliorated
if a singlet neutrino is added which couples as $\lambda^s \ell^c \phi \nu_s$
to the right-handed neutrino.   With only this and the above Yukawa terms,
there exists a massless neutrino state.  A small Majorana mass $m_s$ for the 
singlet produces a light neutrino with mass 
$m_s (\lambda^{\nu} h / \lambda^s \phi)^2$.  This gives a symmetry reason
for the existence of a small quantity -- the fact that non-zero $m_s$ 
violates fermion number.  This mechanism becomes more important for
the case described below.

The resulting model in the IR is simply the SM 
in which all right-handed fermions are composite with a ``pion decay
constant'' of $M \sim 3-4$ TeV and in which $\sint$ is acurately predicted
to order a few percent.  

\noindent{\bf Theoretical Uncertainty:}
The main contributors to the uncertainty in the prediction of $\sint$ 
come from the, in principle, uncorrelated values of the $SU(2)_{\tilde L}\times U(1)_X$
gauge couplings.  As discussed in a previous paper \cite{Dimopoulos:2002mv}, 
these contributions are of order a few percent over a large region of
parameter space in which their couplings are greater than unity.  Another
source is the unknown value of the $SU(3)_W$-breaking scale.  If we assume
it is in the few TeV range, motivated by theories which explain the
hierarchy problem with a low cutoff, the uncertainty is again reduced to
the few percent level.

The corrections to the value of the weak mixing angle from strong dynamics 
can be estimated by NDA and come in the form of operators such as
\begin{equation}
\frac{\Lambda^4}{g^2} \left[ U \frac{g_i^2 F^{\mu\nu} F_{\mu\nu}}{\Lambda^4} \right]
\end{equation}
where the gauge fields are $SU(3)_W$ or $SU(2)_L \times U(1)_X$ fields,
$g_i$ are their gauge couplings and 
$U$ breaks the gauge symmetry.  From this operator, we estimate fractional 
contributions to $\sint$ to be of order $(\alpha_i/4\pi ) / g_W^2 \sim few \%$.
Thus, although this is a strongly coupled theory, a measured
observable can be accurately predicted.

\section{Variations}
One obvious question is why not take the minimal module of 
$SU(3)_W \times SU(2)_{\tilde L} \times U(1)_{\tilde Y}$ and 
let $SU(2)_{\tilde L}$ get
strong.  This model does in fact support complementarity and it can be 
shown via 't Hooft anomaly matching that the gauge symmetry {\it must}
be broken in the IR.  The low energy theory is just the SM with composite
left-handed fermions.  In addition, there is one less theoretical uncertainty
as the $SU(2)_{\tilde L}$ coupling has been removed and only the extra $U(1)_{\tilde Y}$
remains.

However, this model is ruled out.  If we make the very reasonable restriction
that the $U(1)$ Landau pole is at least a factor of $4\pi$ larger than
the breaking scale $M$, then the coupling $\gtp < 2$.  These values of
the coupling decrease the value of $\sint$ at the breaking scale $M$ thus
requiring a smaller $M$ to reproduce the measured result at $\Mz$.
For $\gtp = 2$, the measured value of $\sint$ predicts a breaking scale
of 1.1 TeV.  The bounds on four-fermion operators from atomic parity 
violation measurements require $M > 3-4$ TeV, with a slightly weaker 
bound coming from other operators.  A more concrete number comes from 
LEP bounds on a SM-like $Z'$ gauge boson.  The bound is from the elecroweak 
fit and is about $M_{z'}>900$ GeV.  This limit, though, comes from virtual 
effects and so the coupling-independent bound on the $U(1)$-breaking 
scale is 900 GeV/$g_z \sim 2$ TeV.

These limits can be avoided by noticing that the leptons make up complete
$SU(3)_W$ triplets.  If we introduce three generations of vector-like
leptons in triplet representations, $L_i$ and ${\bar L_i}$ with a 
technically natural small mass term and promote 
$\Sigma$ to the $(3,2,1/2) + (3,1,-1)$
representation, then the couplings $y {\bar L_i} \Sigma (\ell,e^c)$ 
produces a mass for the leptons in the $SU(2)\times U(1)$ sector.  
The SM leptons now do not feel the strong dynamics or the coupling
to a single $Z'$.  Bounds on charged extra gauge bosons from LEP put
a limit on $M$ of about 1 TeV due to the leptons coupling to the rest
of the $SU(3)$ gauge multiplet.  The strongest bound, however, comes
from non-observation of muonium-antimuonium conversion which puts a
limit of $M>1.4$ TeV and $M>850$ GeV at 90\% and 95\% C.L. respectively
\cite{Willmann:1998gd}.  Versions of the model where only one generation
lives in a triplet may avoid these bounds.  Four quark operators also 
put a bound on the breaking scale at $\sim 1$ TeV.

Another possible model is to put $SU(2)_L\times U(1)_{\tilde Y}$ into
the Pati-Salam group \cite{Pati:1974yy}.  The full gauge group is 
$SU(3)_W \times SU(4)_c \times SU(2)_L \times SU(2)_R$, a semi-simple
group in which case charge quantization is guaranteed.  The fermions 
are three generations of $Q: (1,4,2,1)$, $Q^c: (1,{\bar 4},1,2)$ and
$\nu_s: (1,1,1,1)$.  These multiplets become, after Higgsing to the 
SM, the normal fermion content plus three generations of right-handed
and sterile neutrinos. 

The scalars are $\Sigma: (3,1,2,2)$, $h: (1,1,2,2)$ and $\phi: (1,4,1,2)$,
where $\phi$ is responsible for breaking Pati-Salam to the SM-like groups.
The relevant global symmetries are $SU(12)_f \times SU(12)_s \times U(1)_s$
analogous to those in the left-right model.  We postulate the existence of
condensates similar to Eqs. (\ref{condensate1}) and (\ref{condensate2}) 
which break
the global symmetries to $SU(12)_f \times SU(10)_s \times U(1)_s^2$.  
Again the SM is the IR field content (with extra singlet neutrinos) and
the set of goldstones contains the Higgs scalar.

A new constraint due to the Pati-Salam group is on the Yukawa couplings.
The couplings $\lambda^u Q h Q^c$ and $\lambda^d Q h^{\dagger} Q^c$ will
eventually give Dirac masses to all the fermions in the IR, including
the neutrinos.  The neutrino masses are equal to the up-type quark masses
with these Yukawas alone.  Again, as in the left-right model, we
add the coupling $\lambda^s Q^c \phi \nu_s$.  If we don't add Majorana
masses for the singlets $\nu_s$, then the IR theory has a set of left-handed
massless neutrinos.  A small Majorana mass $m_s$ produces a small Majorana
mass for the low-energy neutrinos of size $\sim m_s (m_q/M)^2$.

The down-type quarks and charged leptons also have equal Yukawa couplings
in the UV.  However, at the scale $M \sim few$ TeV, assuming just the SM 
particle content up to that scale, the bottom Yukawa is expected to be
about 50\% larger than that of the $\tau$.  This discrepancy can be understood
in the case when the theory has a low cutoff.  Non-renormalizable operators
of the form:
\begin{equation}
\frac{\lambda'}{\Lambda^2} Q^c \phi \phi^{\dagger} h^{\dagger} Q
\end{equation}
contribute to Yukawa couplings at the Pati-Salam-breaking scale $M$.
If the $SU(4)_c$ indices on the $\phi\,$s are contracted with the $Q\,$s,
then this operator only contributes to the leptons.  Otherwise, when
$Q\,$s and $\phi\,$s are contracted with themselves, the contribution
is quark-lepton universal.  If the cutoff $\Lambda$ is around 30 TeV,
these contributions can be as large as $10^{-2}$, easily enough to 
explain the dicrepancy.  As complementarity suggests, this should
be true in the strongly coupled picture and we leave this as an exercize
for the reader.

Besides charge quantization, a nice feature of this model is the reduction
of theoretical inputs.  Since the gauge coupling of $SU(2)_R$ gets strong
and the $SU(4)_c$ coupling is fixed by the measured value of the QCD coupling,
only the $SU(2)_L$ can be a significant unknown contribution to $\sint$ at
the scale $M$.  There is a large region in this coupling's parameter space
for which the contribution to $\sint$ is order a few percent.  This can be seen
by looking at the right-most boundary of the plot in Figure 2 of 
\cite{Dimopoulos:2002mv}.

The examples of this section have fundamental scalars whose masses
are quadratically sensitive to the cutoff. One option for avoiding this is
a $\sim$ TeV-scale cutoff. Others are the possibility that the scalars are 
composite, or TeV-scale supersymmetry.

\section{Experimental Signatures and Bounds}
The models presented here have two ingredients: those
responsible for the prediction of the weak mixing angle
(such as SU(3)), and the strong $SU(2)_R$ sector. The
experimental consequences associated with the first are the
presence of a weakly coupled SU(3), an $SU(2)\times U(1)$
with intermediate-strength coupling, and $\Sigma$ which
bridges the two. These have already been introduced in
reference \cite{Dimopoulos:2002mv}. The new physics introduced in this
paper is the $SU(2)_R$ force, which becomes strong at a few
TeV and leads to composite right handed particles.

The main experimental bounds to compositeness come from
limits on 4-fermi interactions. Parametrizing their
coefficients as $1\over M_c^2$, we obtain the following
limits to $M_c$: Atomic parity violation limits the
coefficient of the operator llqq, $M_c \geq 4$ TeV. The
strongest bound on the four-lepton operator implies $M_c
\geq 3$ TeV, whereas the limit on the four-quark operator gives
$M_c \geq 1$ TeV (assuming left-left and right-right bounds are
about equal). Naive dimensional analysis gives
coefficients of the four-fermion composite operators that
are of order $1\over M^2$, where M is the analogue of
$f_{\pi}$, leading to the identification $M\sim M_c$. This
shows that these constraints are consistent with the range
of $M$ suggested by the weak mixing angle as well as the
hierarchy problem.  

In models where the leptons are in $SU(3)$ triplets they couple
to doubly-charged gauge bosons and therefore are most sensitively
probed by experiments which look for muonium-antimuonium
conversion ($\mu^+ e^- \rightarrow \mu^- e^+$).  The current limits
on V-A and V+A interactions are \cite{Willmann:1998gd}
\begin{equation}
G_{\rm M {\overline M}} \leq 3.0 \times 10^{-3} \, {\rm G_F} \:\: (90\% {\rm C.L.})
\end{equation}
These bounds already cover an interesting regions in parameter space.
If higher muonium production efficiency can be achieved, the entire
parameter space of most models could be probed.

Perhaps the most exciting signature of the $SU(2)_R$ force
comes from its instantons. Since $SU(2)_R$ gets strong, the
instanton amplitude is large -- just as in QCD. It emits
twelve fermions (nine quarks and three leptons), one for
each right-handed doublet. At the LHC this will result in
spectacular events with many jets and three leptons. The
$SU(2)_R$ instantons can also mediate $\Delta B=\Delta L=3$
nuclear transitions, such as the decay of tritium or
helium-3 to three leptons. The strongest bound
comes from the decay of oxygen into a state which contains
thirteen nucleons and three leptons; this is tested 
in the Super-Kamiokande water-stability experiment.
Since the instanton is a dimension 18 operator, the rate is
supressed by 28 powers of M and is adequately small.

The simplest theory with composite right-handed particles
is the three-family $SU(3)_c \times SU(2)_L \times SU(2)_R
\times U(1)_x$ model, with $SU(2)_R $ getting strong near a
TeV. This model does not have the new $SU(3)_W$ symmetry
(so, it does not predict $\sint$), nevertheless it has all
the signatures associated with composite right-handed
particles
--including $SU(2)_R$ instantons. The model
is complementary, and it is straightforward to verify 't
Hooft's matching conditions.

{\bf Acknowledgments:} We happily thank David B. Kaplan, Ann Nelson and
Leonard Susskind, and Neal Weiner for helpful conversations and comments.
This work is supported by NSF grant PHY-9870115 and DOE grant
DE-AC03-76SF00515.

\bibliography{strong}

\providecommand{\href}[2]{#2}\begingroup\raggedright\begin{thebibliography}{10}

\bibitem{Dimopoulos:1981zb}
S.~Dimopoulos and H.~Georgi, {\it Softly broken supersymmetry and su(5)},  {\em
  Nucl. Phys.} {\bf B193} (1981) 150.

\bibitem{Georgi:1974sy}
H.~Georgi and S.~L. Glashow, {\it Unity of all elementary particle forces},
  {\em Phys. Rev. Lett.} {\bf 32} (1974) 438--441.

\bibitem{Georgi:1974yf}
H.~Georgi, H.~R. Quinn, and S.~Weinberg, {\it Hierarchy of interactions in
  unified gauge theories},  {\em Phys. Rev. Lett.} {\bf 33} (1974) 451--454.

\bibitem{Dimopoulos:2002mv}
S.~Dimopoulos and D.~E. Kaplan, {\it The weak mixing angle from an su(3)
  symmetry at a tev},
  \href{http://xxx.lanl.gov/abs/http://arXiv.org/abs/hep-ph/0201148}{{\tt
  http://arXiv.org/abs/hep-ph/0201148}}.

\bibitem{Dimopoulos:1980hn}
S.~Dimopoulos, S.~Raby, and L.~Susskind, {\it Light composite fermions},  {\em
  Nucl. Phys.} {\bf B173} (1980) 208--228.

\bibitem{tHooft}
G.~'t~Hooft, {\it Recent developments in gauge theories},  {\em eds. G. 't
  Hooft, et al} {\bf Plenum Press, N.Y.} (1980).

\bibitem{Fradkin:1979dv}
E.~H. Fradkin and S.~H. Shenker, {\it Phase diagrams of lattice gauge theories
  with higgs fields},  {\em Phys. Rev.} {\bf D19} (1979) 3682.

\bibitem{Weinberg:1972nd}
S.~Weinberg, {\it Mixing angle in renormalizable theories of weak and
  electromagnetic interactions},  {\em Phys. Rev.} {\bf D5} (1972) 1962--1967.

\bibitem{Coleman:1982yg}
S.~R. Coleman and B.~Grossman, {\it 't hooft's consistency condition as a
  consequence of analyticity and unitarity},  {\em Nucl. Phys.} {\bf B203}
  (1982) 205.

\bibitem{Frishman:1981dq}
Y.~Frishman, A.~Schwimmer, T.~Banks, and S.~Yankielowicz, {\it The axial
  anomaly and the bound state spectrum in confining theories},  {\em Nucl.
  Phys.} {\bf B177} (1981) 157.

\bibitem{Weinberg:1979kz}
S.~Weinberg, {\it Phenomenological lagrangians},  {\em Physica} {\bf A96}
  (1979) 327.

\bibitem{Manohar:1984md}
A.~Manohar and H.~Georgi, {\it Chiral quarks and the nonrelativistic quark
  model},  {\em Nucl. Phys.} {\bf B234} (1984) 189.

\bibitem{weak}
H.~Georgi, {\it Weak interactions and modern particle theory},  {\em
  Benjamin/Cummings Publishing Company, Inc.} (1984).

\bibitem{Willmann:1998gd}
L.~Willmann {\em et.~al.}, {\it New bounds from searching for muonium to
  antimuonium conversion},  {\em Phys. Rev. Lett.} {\bf 82} (1999) 49--52,
  [\href{http://xxx.lanl.gov/abs/http://arXiv.org/abs/hep-ex/9807011}{{\tt
  http://arXiv.org/abs/hep-ex/9807011}}].

\bibitem{Pati:1974yy}
J.~C. Pati and A.~Salam, {\it Lepton number as the fourth color},  {\em Phys.
  Rev.} {\bf D10} (1974) 275--289.

\end{thebibliography}\endgroup

\bibliographystyle{JHEP}

\end{document}